\RequirePackage{lineno}
\documentclass[fleqn,10pt]{wlscirep}

\title{Detecting and modelling  delayed density-dependence in abundance 
time series of a small mammal ({\it Didelphis aurita})}

\author[1,*]{E. Brigatti}
\author[2]{M. V. Vieira}
\author[3]{M. Kajin}
\author[4]{P. J. A. L. de Almeida}
\author[5]{M. A. de Menezes}
\author[2]{R. Cerqueira} 
\affil[1]{Instituto de F\'{\i}sica, Universidade Federal do Rio de Janeiro, 
Av. Athos da Silveira Ramos, 149,
Cidade Universit\'aria, 21941-972, Rio de Janeiro, RJ, Brasil.}
\affil[2]{Laborat\'orio de Vertebrados, Instituto de Biologia, Universidade Federal do Rio de Janeiro. Caixa Postal 68020, 21941-590, Rio de Janeiro, RJ, Brasil.}
\affil[3]{Laborat\'orio de Ecologia de Mam\'{\i}feros, Departamento de Ecologia, Instituto de Biologia Roberto Alcantara Gomes, Universidade do Estado do Rio de Janeiro, Campus Maracan\~a, 20550-900, Rio de Janeiro, RJ, Brasil.}
\affil[4]{Coordena\c c\~ao de Matem\'atica Aplicada, Laborat\'orio Nacional de Computa\c c\~ao Cient\'{\i}fica, 25651-075, Petr\'opolis, RJ, Brasil.}
\affil[5]{Instituto de F\'{\i}sica, Universidade Federal Fluminense, 
 Campus da Praia Vermelha, 24210-340, Niter\'oi, RJ, Brasil.}
\affil[*]{edgardo@if.ufrj.br}

\begin{abstract} 

We study the  population size time series of a Neotropical small mammal with the intent of detecting and modelling population regulation processes 
 generated by density-dependent factors and their possible delayed effects.  
The application of analysis tools based on principles of statistical generality 
are nowadays a common practice for describing these phenomena, 
but, in general, they are more capable of generating clear diagnosis rather than
granting  valuable modelling.
For this reason, in our approach, 
we detect the principal temporal structures on the bases of 
different correlation measures, and from these results we build an ad-hoc minimalist autoregressive model that incorporates the main drivers of the dynamics. 
Surprisingly our model is capable of reproducing very well the time patterns of the empirical series and, for the first time, clearly outlines the importance of the time of attaining sexual maturity as a central temporal scale for the dynamics of this species. In fact, an important advantage of this analysis scheme is that all the model parameters are directly biologically interpretable and potentially measurable, allowing a consistency check between model outputs and independent measurements.

\end{abstract} 

\begin{document} 

\flushbottom
\maketitle


\section*{Introduction}

One of the main objectives in the field of population dynamics is to determine the extent of deterministic vs. stochastic forces in time series of abundance and population parameters.
The deterministic share of the observed fluctuations is usually assigned to nonlinear density-dependent processes, which create regulatory and stabilising forces \cite{Berryman2008}.
Different theoretical and modelling frameworks have been used through the history of population dynamics, but time series analysis and autoregressive models are a frequent and natural choice, as the population size in the future is related to the population size in the past \cite{Lande2003,HendersonMagurran2014,Sibly2005}.
A variety of time series analysis methods have been used in population dynamics to diagnose their structure and density dependence \cite{Turchin2003}, particularly successful in the analysis of empirical data of long lived taxa such as mammals\cite{grenfell,Fryxell1998,FergusonPonciano2015}. 
The approach proposed by Royama \cite{Royama92} combines diagnostic tools with the use of phenomenological models, and has increased the predictive power and understanding of the dynamics of intensively studied systems \cite{Lima2008,BerrymanLima2006,Boonstra}. 
In general, for sufficiently long time series, it is possible to use conventional autoregressive models or to apply other analogous methods which aim to determine a 
clear differentiation between deterministic and random components \cite{Dennis1994,EllnerTurchin1995}. 
The most common approach \cite{Royama92,Bjornstad} uses a linear autoregressive 
model of order $k$ , 
which relates the logarithm of the population abundance ($y_t$) at different time steps:
\begin{linenomath}
\begin{equation}
y_t=\beta_0+\sum_{i=1}^k \beta_i y_{t-i} + \epsilon_t.
\label{equation1}
\end{equation}
\end{linenomath}

As $\epsilon_t$ represent a random component, this is an ordinary $AR(k)$ model,
where the  k-value can be determined implementing an AIC-based selection.
This approach is equivalent to fitting the population growth rate 
and it is efficient in determining density dependence. 
Considering that the correct estimation of the order of the model can be critical, 
that the real process can be highly nonlinear \cite{Royama92, Turchin1993}, 
and that a direct connection between parameters $\beta_i$
and demographic estimated quantities is difficult,
in general, the most appropriate role 
of this method is diagnosis rather than modelling \cite{Berryman01}. 
Finally, taking into account that the random component included 
in population size time series 
can be difficult to characterise \cite{ahrestani}, 
and that time series of the necessary length are not so common, it may be reasonable to take the reverse path. 
Instead of trying to diagnose and project the exact time series behaviour based on a general autoregressive model of a given order, it can be more informative to make inferences about its statistical properties and patterns of variance, 
and use these results to elaborate a specific and realistic dynamical model. 
Finally, the parameters of this particular model may be estimated by a standard autoregressive procedure.

Here in we use diagnostic tools to characterise the structure of a 13-year time series of population size estimates of a Neotropical marsupial, the black-eared opossum, {\it Didelphis aurita} (Wied-Neuwied 1826), and develop a simple model that successfully reproduces its population dynamics. First, we determine periodicity, autocorrelations structures and the nature of potentially density dependent regulation. 
Second, we use this diagnosis to build a simple model, incorporating the characteristics of the series. 
Third, we estimate the parameters of such a model.
Finally, we compare the observed series with the ones generated by the model as an empirical test of model fit to the data, and of the whole procedure of model development. 
Models generated by this procedure can be simpler compared to some aspects of previous approaches, and yet still incorporate the main drivers of the dynamics. The simple model developed for {\it D. aurita} is capable of reproducing the time series dynamic behaviour and all their parameters are biologically interpretable and measurable in term of basic demographic quantities.

\section*{Materials and methods}

We analysed a 13-year time series of local population size estimates of black-eared opossums.
The data set originated from the long-term small mammal population-monitoring program performed by Laborat\'orio de Vertebrados, Universidade Federal do Rio de Janeiro, within the Parque Nacional da Serra dos \'Org\~aos, state of Rio de Janeiro ($22\,^{\circ}$ 28' 28'' S, $42\,^{\circ}$ 59' 86'' W, details in \cite{datadidelphis}). 
The local population sizes were estimated using program MARK \cite{CoochWhite1994}. To minimise census error, accurate population size estimates were obtained in a two-step modelling procedure: we first tested for the existence of heterogeneity in capture probabilities, then, having evidence of the existence of the latter, we used the appropriate estimator, namely, the Jackknife estimator, to obtain population size estimates. Different models were fitted to the data, where the candidate model set comprised models including temporal variation in all real parameters and all nested models within this global model. The model that was best supported by the data was chosen based on the maximum likelihood principle (using QAICc index, corrected for possible data over-dispersion) \cite{Burnham}. 

\begin{figure}[ht]
\begin{center}
\includegraphics[width=0.75\textwidth, angle=0]{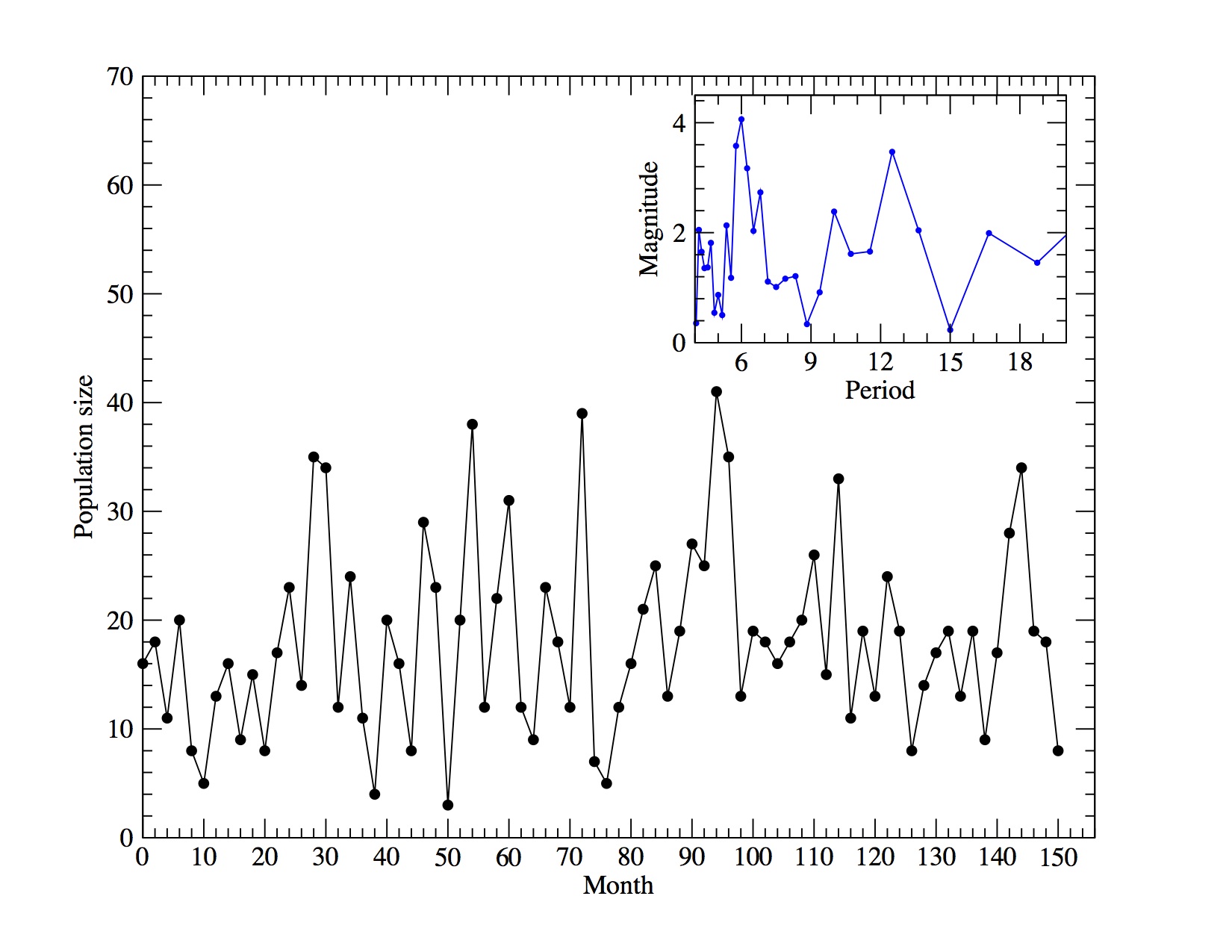}
\end{center}
\caption{\small  The time series used, comprised of bimonthly estimated local population sizes of a {\it Didelphis aurita} population in the Brazilian Atlantic forest, State of Rio de Janeiro, from April 1997 to October 2009. In the inset, the magnitude of the Discrete Fourier Transform of the data against the period.
}

\label{fig_time}
\end{figure}

The time series used in the analyses ($X_{t=1,2,...,76}$) represents the estimated bimonthly local population sizes from April 1997 to October 2009 (see Figure \ref{fig_time}). 
Most studies of mammal populations analyse abundance counts or estimates using annual data \cite{aanes,murua,lima06}, and rarely populations are analysed on a finer time scale, using seasonal \cite{lima01} or monthly abundances \cite{leirs}. 
Annual abundances are more commonly used when the aim is to describe long term behaviours, the organism has a defined and limited reproductive season within the year, or simply because data on a finer time scale are not available. 
In our case, reproduction of {\it D. aurita} occurs during a large portion of the year with no single reproductive peak \cite{RademakerCerqueira}, having two or up to three litters within a year \cite{datadidelphis,maya}. 
Moreover, population size fluctuations may not be related exclusively to reproductive events: immigrations to and emigrations from the area may occur \cite{Oli01}, modifying the local population size on a time scale that is not season or year-related. 
At the same time, both of these processes may be density dependent. 
Therefore, the appropriate temporal scale to investigate changes in abundance and regulatory processes must be thinner than a year. This suit of life history traits is common to other populations of neotropical marsupials \cite{gentile}, hence seasonal or monthly time scales are likely the most appropriate.  \\

As a first step, we try to identify basic patterns that clearly deviate from pure random behaviour. 
These basic patterns should be capable of summarising statistical and dynamical behaviours, with attention to the broad essential structures rather than the specific details. 
We emphasise that our time series is still not long enough to consider particular quantitative numerical outputs of the statistical analyses, but rather we limited our considerations to the qualitative behaviour delineated by our examination.

\begin{figure}[ht]
\begin{center}
\includegraphics[width=0.75\textwidth, angle=0]{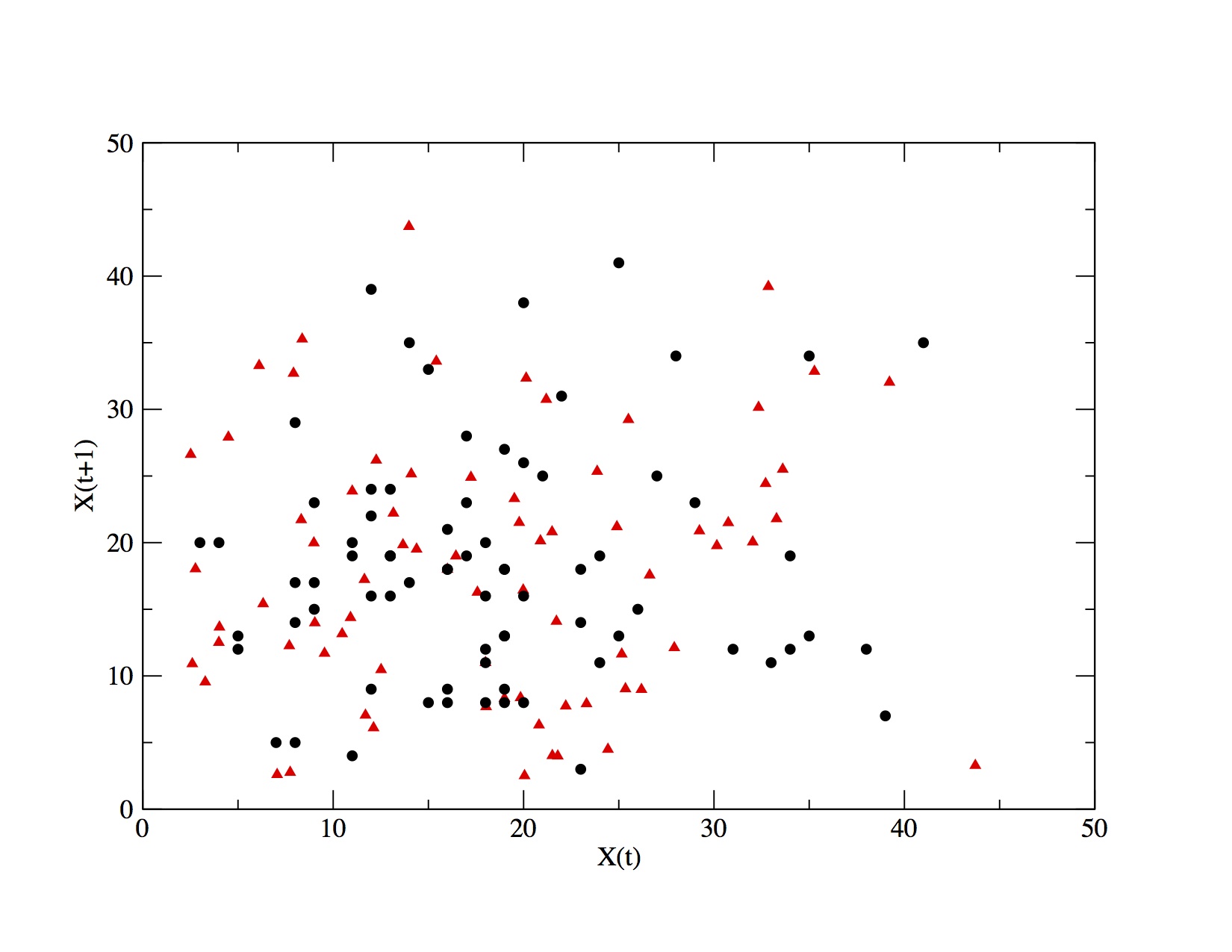}
\end{center}
\caption{\small Comparison of the observed $X(t+1)$ versus $X(t)$ (black circles) with the iterates of the best-fit model parametrised as in Table \ref{table} with added noise (red triangles). 
}
\label{fig_map}
\end{figure}

\begin{figure}[ht]
\begin{center}
\includegraphics[width=0.75\textwidth, angle=0]{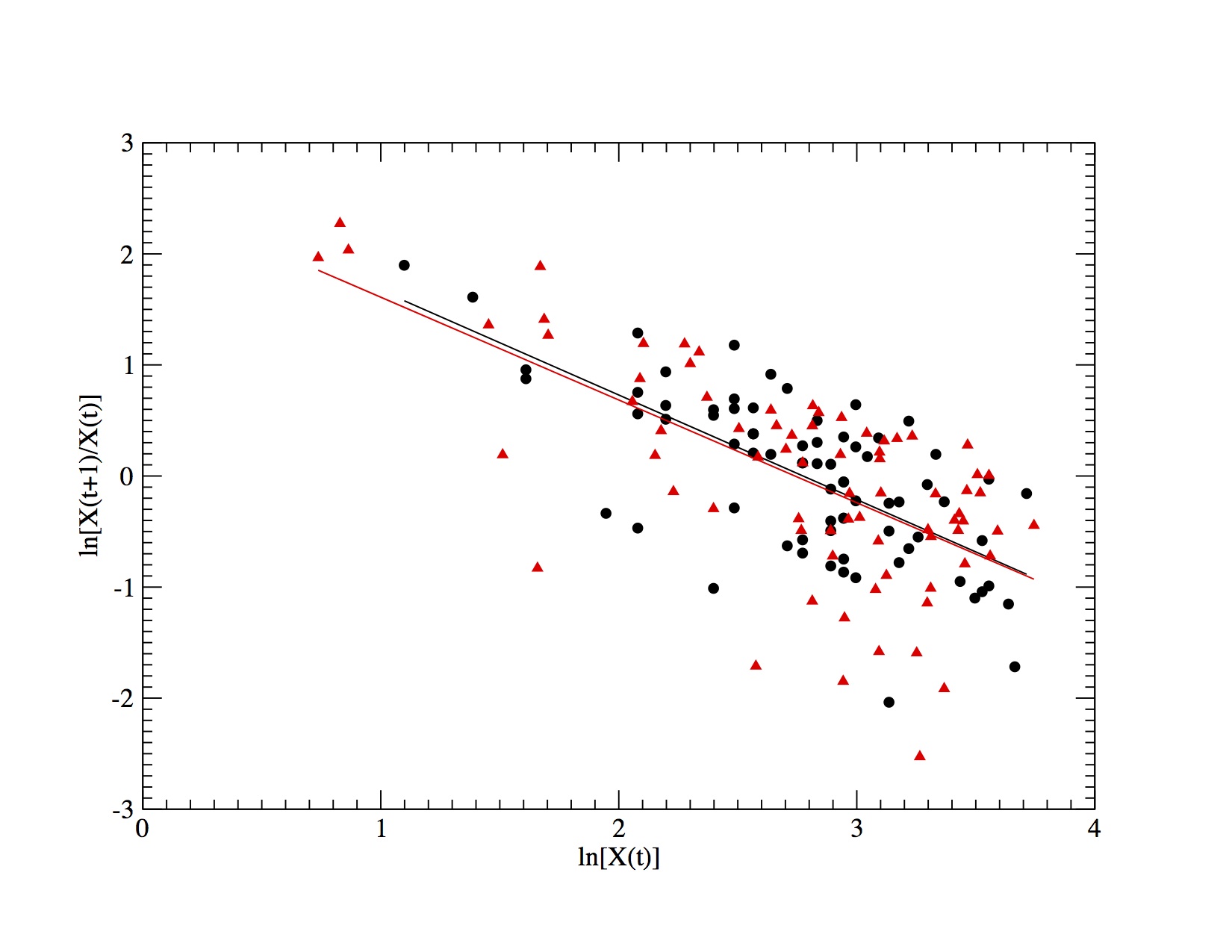}
\end{center}
\caption{\small Observed log growth rate versus log abundance (black circles), with iterates of the best-fit model with the parameters of Table \ref{table} and the added noise (red triangles). 
The black line represents the linear fit for the real data: $\log{\frac{X(t+1)}{X(t)}}=2.61-0.94\times\log{X(t)}$, and the red one
for the simulated data: $\log{\frac{X(t+1)}{X(t)}}=2.53-0.93\times\log{X(t)}$.
}
\label{fig_Growth}
\end{figure}

The standard plot of $X(t+1)$ against $X(t)$ shows a uniform
distribution of points suggesting that no threshold-like
behaviour or elementary correlations exist  in our data (see Figure \ref{fig_map}).
The plot of the logarithm of the population growth rate ($\log{\frac{X(t+1)}{X(t)}}$)
as a function of the logarithm of the population number ($\log{X(t)}$) shows
that the growth rate clearly declines as population size increases (see Figure \ref{fig_Growth}). The relationship seems also characterised by a
larger variance at high population sizes but it  
is difficult to identify this effect because of the small sample sizes at low abundances.
This behaviour in the variance can be generated by involved interactions between process errors (biotic and abiotic) and observation errors. 
Estimation of these different sources of error is not possible, and out the scope of this study.

\begin{figure}[ht]
\begin{center}
\includegraphics[width=0.75\textwidth, angle=0]{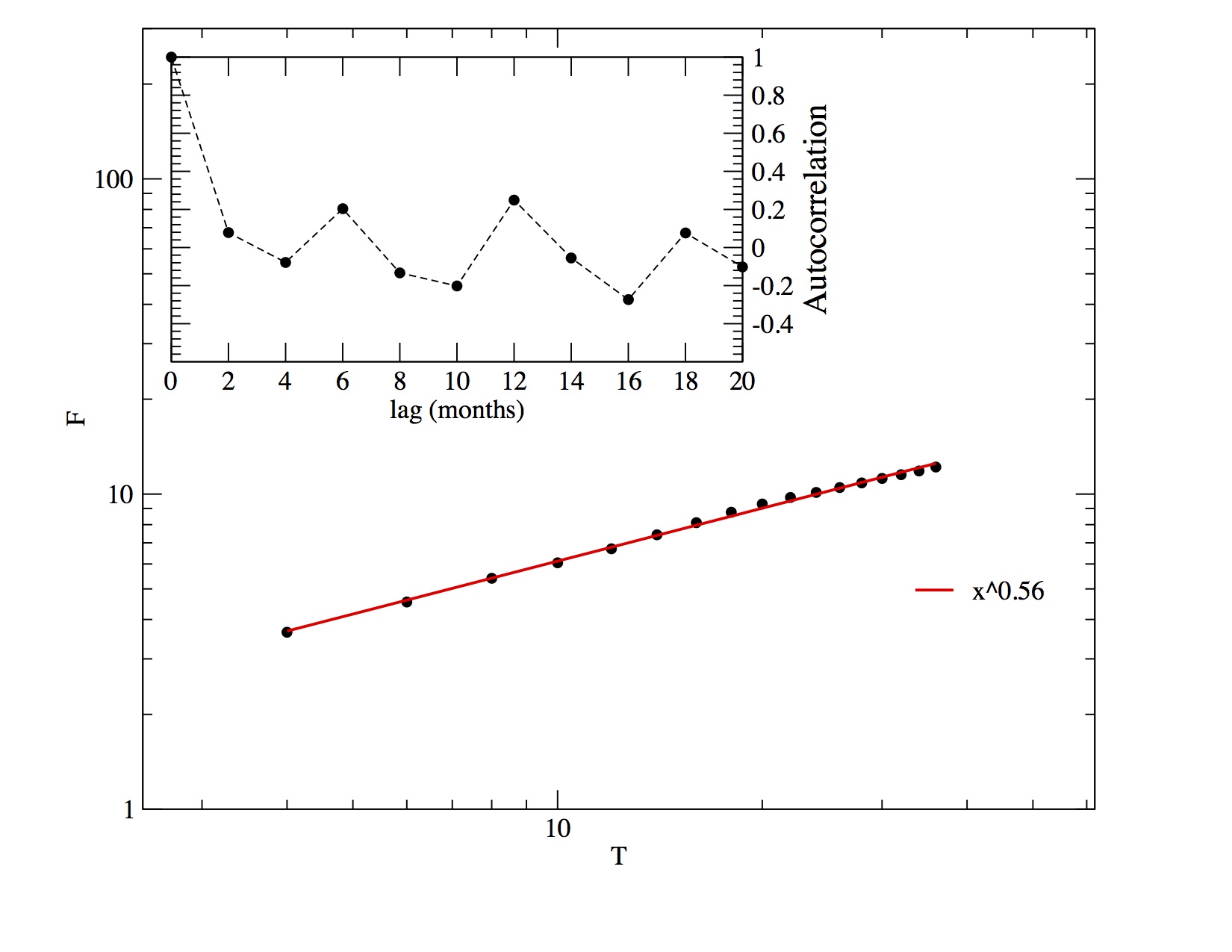}
\end{center}
\caption{\small DFA analysis of the abundance time series.
 The red line represents the fit: $F(T)= 1.7 \times T^{0.56\pm0.01}$.
In the inset, the autocorrelation function of the same dataset.
}
\label{fig_DFA}
\end{figure}

We performed a spectral analysis of the data, 
which does not present any robust or clear
periodicity, except for some more prominent modes in correspondence of 
the 6-month time
lag and its multiples (see inset of Figure \ref{fig_time}).
We continued our inspection with the study of the autocorrelation function. 
Its behaviour reveals an interesting periodical structure (inset of Figure \ref{fig_DFA}), where positive autocorrelations were found for 6-month time lag and its multiples. The strongest negative autocorrelation values appear always four months after the positive peaks.  

We explored the long term behaviour of correlations
looking at the scaling of the time-series fluctuations at different time lags.
Power-law scalings in population fluctuations, which display a clear deviation from pure random behaviours, are well known fact reported in the literature \cite{marquet}.
The direct inspection of autocorrelations is usually 
not appropriate because of noise effects and  because of possible non-stationarities \cite{peng}. 
In contrast, a robust and general method for the characterisation of the
fluctuations behaviour at different scales is the detrended 
fluctuation analysis (DFA) \cite{peng}. 
It measures the second moment-fluctuation $F$ of the detrended time series for different 
time windows of size $T$, obtaining $F=F(T)$.
The scaling behaviour is characterised by estimating the exponent $\alpha$, a generalisation of the Hurst exponent, which is extracted from the relation $F(T)\propto T^{\alpha}$.
A straightforward way for its evaluation is a linear fit of the log-log
graph of $F(T)$ as a function of $T$ (Figure \ref{fig_DFA}).
For our time series we obtain $\alpha=0.56\pm0.01$ , which is significantly different 
from the case of uncorrelated, white noise series, characterised by  $\alpha=1/2$.
Considering that the exponent is not very far from 1/2 and that the regression is realised 
on less than two decades, we are lead to interpret these outcomes 
just as a detection of positive autocorrelations rather than a claim about the 
self-similarity of the time series. 
For this reason we use the DFA method for clearly identifying positive correlations,
not for conclusively assessing their power-law scaling \cite{morariu, marcio}.

We can summarise the principal results of this analysis
outlining  two basic patterns
present in our data.
First, a clear trend in the growth rate which declines for large population sizes \cite{grenfell}.
This effect suggests the presence of some type of density dependent
regulation in the population dynamics.
Second, the autocorrelation analysis sketch out 
a positive correlation for lag multiples of six months,
followed by a negative one four months later.
These results are corroborated by the DFA analysis, where 
we can outline a deviation from an uncorrelated, 
white noise time series ($\alpha>1/2$), which indicates some type of 
long-memory process that generates  correlations between the data 
\cite{kante}.
These outcomes 
indicate a possible time delay in the density effect on natality
and suggest the use of some delayed-recruitment
model for the replication of the time series. 
\\

Starting from these results we assemble one of the simplest
autoregressive models which could account for such 
basic patterns. 
In fact, as we study a specific population, it is 
more appropriate to build in its specific
mechanisms and patterns, rather than to bother with claims of statistical generality.
We begin from the simplest recurrence relation
$x_{t}= ax_{t-1}$,
where $x_{t}$ denotes the population size at time $t$.
Each period $t$ represents two months, 
which correspond to the time scale of field data collection.
A recruitment term must be added, which can account 
for the detected structure in the correlation of the time series, and 
takes into consideration that sexual maturity 
in this population takes place only 6 months after birth \cite{maya}.
For these reasons, we add a second term $G(x_{t-\tau})$ 
where $x_{t-\tau}$ is the adult breeding population, which is selected with a delay 
of $\tau$ periods. 
An analogous model has been used for studying baleen whale 
populations \cite{allen,clark}.
The final strategy is to make the minimalistic 
assumption about the form of $G(x_{t-\tau})$
which could describe the outlined density dependent 
growth rate. A common one is a quadratic polynomial,
which leads to the following autoregressive model:

\begin{linenomath}
\begin{equation}
x_{t}= ax_{t-1}+bx_{t-\tau}+cx^2_{t-\tau}+\epsilon_{t}.
\label{eq_model}
\end{equation}
\end{linenomath}

Here $\epsilon_{t}$ stands for some stochastic term  affecting the population abundance 
$x_{t}$ and, as a starting approximation, we model it with a normally distributed noise $N(0,\sigma^2)$.
The autoregressive coefficients can be obtained by
least-squares estimates: we minimise the squared difference 
between the modelled time series and the observed one \cite{victor}. 
We based our model on the observed value of the population abundance $X_t$, 
but the use of the $\log{X_t}$ generates very similar results.

Finally, we must determine the value of $\tau$. This is attained identifying which is best at 
predicting the observed data. 
The accuracy of our prediction is measured using the following quantity  \cite{turchin}:
\begin{linenomath}
\begin{equation}
R^2=1-\frac{\sum^{76}_{i=1+\tau}(X_i-\widehat{X_i})^2}{\sum^{76}_{i=1+\tau}(X_i-\bar{X})^2}
\end{equation}
\end{linenomath}
where $\widehat{X_i}$ are the predicted values, $X_i$ are the observed ones and $\bar{X}$ is their mean value.
$R^2=1$ represents a perfect prediction and accuracy reduces with decreasing in $R^2$ value. 
We draw attention to the fact that negative values are a possible output since we
are not simply fitting our values with a regression, but we are trying to 
predict future values with an autoregression \cite{turchin}.
In particular, negative values indicate that prediction is even worse than using the mean of the series.
This quantity can be used for determining the best $\tau$ value only because the number 
of model parameters is fixed, and it does not change with the value of  $\tau$, like in the 
approach presented in equation \ref{equation1}.
Finally, an advantage in the use of $R^2$ over a likelihood based approach is that 
it can be compared over different sample sizes  
(data must not be thrown out to directly compare different  $\tau$ values). 

We can summarise some positive aspects in the use of this model over 
the standard approach of equation 1.
Our approach is easier because it can account for different lags in the 
delayed density dependence with a fixed number of parameters. 
This is an important point for data having a time scale finer than one year.
Detecting a two-year delay for bimonthly data using the standard approach implies the estimation of thirteen parameters, an impracticable task because of the curse of dimensionality.
In our approach, all model parameters can be related to biological quantities 
that regulate the individual dynamics, and potentially can be measured empirically.
 This allows independent estimation from other demographic studies, important to perform a consistency check of the autoregressive estimation. 
In contrast, in the standard approach, parameters $\beta_i$ are just statistical parameters
for fitting  general delayed density-dependent effects in the population growth rate.
This difference is important if our interest is modelling rather than just detecting
delayed density dependence.
Finally, we model our population at the fundamental level of
abundance and not to the derived level of the growth-rate.

\section*{Results and discussion}

\begin{figure}[ht]
\begin{center}
\includegraphics[width=0.75\textwidth, angle=0]{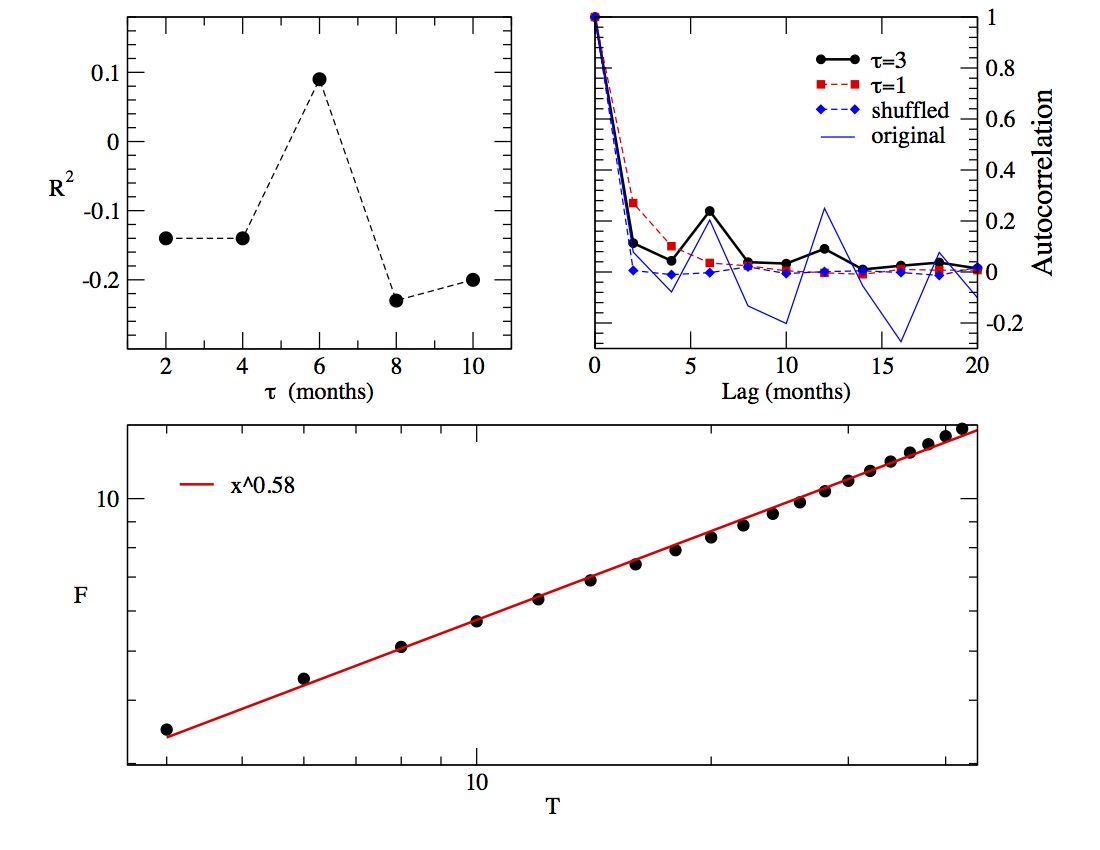}
\end{center}
\caption{\small Top, on the left: the $R^2$ value for different $\tau$. 
Top, on the right: the autocorrelation function of the synthetic series generated by the best fit model,  with added noise, for $\tau=3$ and $\tau=1$.
Blu points are obtained with a random shuffling of the series obtained for $\tau=3$.
From the comparison, we can appreciate the statistical relevance of the autocorrelation values.
Bottom: DFA analysis of the synthetic series generated by the model parametrised as in Table \ref{table} with added noise.
}
\label{fig_DFAbis}
\end{figure}

The analysis presented in Figure \ref{fig_DFAbis} shows that the first positive $R^2$ value is found for $\tau=3$, which corresponds to 6 months.
Temporal delays in population responses can result from the time it takes for many intrinsic (e.g reproductive event, mortality of young) and extrinsic factors (e.g. predator, competition, parasite infection) to result in changes in population parameters \cite{Hudson1998}. These delays are directly related to a number of life-history traits, such as generation time, time of first reproduction, maternal care, reproductive strategy, as well as the species trophic position and the environmental conditions a population faces at a certain moment.
In {\it Didelphis aurita}, a 6-month delay corresponds to the time it takes for the newborn females to reach sexual maturity and start reproducing, also becoming more trappable \cite{maya}. This is the simpler explanation for the positive 6-month delay in our model.
The model present two terms, controlled by parameter $b$ and $c$,
both characterised by a 6-month time delay.
Together they are the responsible for the regulatory process of the population: low population sizes when individuals are born result in $bx_{t-\tau}>cx^2_{t-\tau}$, a positive contribution to the population 6-months latter; larger population sizes when individuals are born quickly results in $bx_{t-\tau}<cx^2_{t-\tau}$, a negative contribution to population size 6-months latter.  Annual growth rates in this population are most sensitive to changes in the survival of pouch young, the newborns of the season \cite{Cerqueira1993,Ferreira2013}. Therefore, the negative density dependence 
 more likely reflects density dependence in survival rates of pouch young. Adult and aged-adult age classes have the relatively highest contributions to fertility \cite{Ferreira2013}, hence more subtle density dependent effects may occur through contributions to fertility by adults and aged adults. However, these would correspond to time delays of 12 months or more as the condition of adults at the start of the breeding season will depend on the conditions of previous seasons, a maternal effect.
 \cite{Beckerman2002}.


The next positive $R^2$ value is naturally found for a multiple of this value,
but not for other $\tau$ values. 
This is the first impressive  consistency check between 
our model outputs and independent measurement of the biological population.
Interestingly, the fact that $R^2<0$ for $\tau=1$ rules out the possibility of using the considered autoregressive model without any delay for successfully describing our data.
No delay in our time series would have no biological support as well.\\

\begin{table}
    \centering
{\small
    \begin{tabular}{|c|c|c|c|c|}
\hline
Parameter & Estimate & Standard Error & $t$ & $P$ \\
\hline
\hline
a & 0.134  & 0.102 & 1.314    &  0.197 \\
\hline
b & 1.514  & 0.243 & 6.230  &  $< 0.001$\\
\hline
c & -0.029  & 0.007 & -4.143   &  $< 0.001$ \\
\hline
\end{tabular}
}
\caption{ Parameter estimates and associated significance levels for the autoregressive model with $\tau=3$.
}
\label{table}
\end{table}

The generation of a simulated time series from the stochastic 
model of equation \ref{eq_model} is straightforward.
We fix the numerical values of the parameters $a,b,c$ and $\sigma^2$  
corresponding to the autoregressive estimation of the best fit model for $\tau=3$
(see Table \ref{table}). Selected the initial conditions $X_1, X_2, X_3$ from the real data,
we obtain a simulated sequence of $X_t$ values.
The simplicity of this approach allows the generations of a number of different trajectories produced by our model which turn out to be a powerful strategy for testing 
our hypothesis. In fact, we can probe if our model well incorporate the  
stochastic mechanism underlying real data and if it can effectively reproduce the 
basic patterns of the observed time series.

First, we can reproduce the general behaviour of the empirical  time series, 
as shown in the scatter plot of $X_{t+1}$ versus $X_t$ of Figure \ref{fig_map}.
More importantly, starting from the synthetic series, we can replicate the 
essential trend of the population growth-rate 
(see Figure \ref{fig_Growth}),
with a clear density-dependent limitation on growth for hight population sizes.
This finding  is very significant. It proves that equation \ref{eq_model} 
can perfectly account for the density dependent effects present in our empirical data:
an approximately  linear dependence of the population growth rate with the log population size.
This result is remarkable because it is obtained modelling the abundance time series and not directly the growth rate, as in standard approaches of the type of equation \ref{equation1}. 
In fact, if we use the model of equation \ref{equation1} we are able to
well reproduce the growth rate behaviour, but,
in this case, this result is quite obvious, as the autoregression is 
a direct fitting procedure over the population growth rate.
All these results are robust and have been tested for different simulation runs and for 
long simulations of up to 10000 time steps.\\

In the following, we examine  the correlations behaviour of the  simulated time series.
For this reason, we generate long simulated series, with at least 1000 events,
which permit robust statistical analysis.
The resulting simulated series present positive correlation values at lag of 6 months and its multiples, a result comparable with the output of the real data analysis (see Figure \ref{fig_DFAbis}). 
The analysis of the time series generated by the model
by means of the DFA method results in an exponent $\alpha$ 
compatible with the one of the real data (see Figure \ref{fig_DFAbis}). 
This fact suggests that the elaborated model 
can capture some aspects of the essential time correlations present in real data.
Only the oscillating behaviour caused by the anti-correlated values that follows the positive peaks are lost in our model.\\ 

One of the most interesting aspect of our autoregressive model is 
that all the parameters can be directly biologically interpreted  
and, potentially, estimated empirically.
The parameter $a$ corresponds to the survival fraction of the population 
that contributes with the population size two months later.
For this reason, its estimated value should be $0<a<1$. 
The parameter $b$ is proportional to the 
fertility, 
and in order to maintain a positive equilibrium population and 
compensate the mortality,
should be $b>1-a$ \cite{clark,murray}.  
The parameter $c$ controls the reduction in fertility
for high populated communities. 
In principle, also this parameter can be related to basic
measurable quantities. \cite{murray}  
It is the responsible for the density-dependent pattern in the population 
growth rate and should be negative.
We can highlight some plausible factors that could be associated to this parameter.
Resource limitation is more likely the main mechanism of density-dependence. 
However, disease cannot be disregarded because of the relatively high load of parasites in animals in the area. 
In addition, high population density may have at least three different effects on future population size through reproduction: (1) prevents active females to reproduce through strong intraspecific competition for food and space, (2) reduces the number of fertilised females by forcing males do disperse due to intraspecific competition, with possible 
modifications in movements strategies \cite{paulo}, and 
(3) forces newborn young to disperse to adjacent areas, 
with a final reduction in local population size estimates.

These constraints limiting the values of model parameters were always respected in our analyses, for all values of $\tau$.
Additional analyses performed using the time series 
generated by different population size estimates (minimum number known alive, ad hoc estimators for each sampling session) confirm these results.
These facts corroborate the robustness of our modelling approach.
In particular, our estimated values for $\tau=3$ satisfy the condition
$1-a<b<3(1-a)$  which states that the model equilibrium solution is locally stable \cite{clark}. 
A final proof of the functionality of our model is the  similarity between our $a$ value
evaluation compared with an independent empirical measurement:
the value estimated by the model corresponds to $a \approx 0.2$, the same found in a demographic study of the same population of opossums \cite{maya}, hardly a coincidence. \\

The possibility of robustly reproducing general qualitative patterns 
with a relatively short time series is a relevant and novel contribution presented here. 
In addition, population size fluctuations of an organism living in an environment as complex as the Atlantic forest produce time series of high variance and noise, adding difficulty to the task.  
Detecting evidence of density-dependent dynamic behaviour is by far not new, but being able to reproduce directly the dynamics of the abundance time series based on biologically realistic and measurable parameters is. 
Our model incorporates the basic demographic characteristics operating in a natural population: the temporal delay due to reproduction peculiarities, fecundity and survival rates, and the effect of density-dependence on reproduction. 




\section*{Acknowledgements}

We are grateful to generations of Laborat\'orio de Vertebrados students that make possible such a long term field work. We also thank A. Marcondes, N. Barros, R. Juazeiro and R. S. Honorato for technical and clerical help. This work has been supported by grants to RC from CAPES, CNPq, FAPERJ, PELD/MCT/CNPq, PROBIO II/MCT/MMA/GEF, and PPBio/CNPQ.






\end{document}